\documentclass[12pt,preprint]{aastex}
\shorttitle{Large-Scale Quasar Jets}
\shortauthors{Stawarz et al.}
\usepackage{natbib}

\begin{document}

\title{On Multiwavelength Emission of Large-Scale Quasar Jets }

\author{\L ukasz Stawarz$^1$, Marek Sikora$^2$, Micha\l \ Ostrowski$^1$ and Mitchell C. Begelman$^3$}
\affil{
$^1$Obserwatorium Astronomiczne, Uniwersytet Jagiello\'nski,\\ 
ul. Orla 171, 30-244 Krak\'{o}w, Poland\\
$^2$Centrum Astronomiczne im. M. Kopernika,\\
ul. Bartycka 18, 00-716 Warszawa, Poland\\
$^3$Joint Institute for Laboratory Astrophysics, University of Colorado,\\
Boulder, CO 80309-0440, USA}
\email{stawarz@oa.uj.edu.pl}

\begin{abstract}

We discuss morphological properties of the large-scale jets in powerful radio sources, which are now observed at radio, optical and X-ray frequencies, in order to determine the origin of their X-ray radiation and the nature of the particle acceleration responsible for their multiwavelength emission. We show that modeling knots of the these objects as stationary regions of energy dissipation within the uniform and continuous jet flow leads to several problems. Such problems are especially pronounced if the observed X-ray emission is due to inverse-Compton scattering of the CMB radiation. We explore another possibility, namely that the knots represent moving and separate portions of the jet matter, with excess kinetic power. We suggest a possible connection between this scenario and the idea of intermittent/highly modulated jet activity. The proposed model can explain some morphological properties of quasar jets --- like high knot-to-interknot brightness contrasts, frequency-independent knot profiles and almost universal extents of the knot regions --- independently of the exact emission mechanism responsible for producing the X-rays. In this context, we consider different possibilities for the production of such X-ray radiation, and discuss the related issue of particle acceleration. We conclude that the appropriate process cannot be designated for certain yet. We suggest, however, that X-ray observations of the jet in 3C 120 (in which intermittent jet activity is quite distinct) already seem to support the synchrotron origin of the X-ray emission, at least in this object.

\end{abstract}


\section{Introduction}

Growing numbers of large-scale jets in powerful extragalactic radio sources are now being observed at optical and X-ray frequencies \citep[see, e.g.,][]{staw04}. New, and in many ways unexpected, results call for a wide discussion of the physical conditions within these objects, and in particular of the processes responsible for accelerating radiating particles to the required ultrarelativistic energies. Until now, few observations had gone beyond the integrated spectral properties of multiwavelength emission from large-scale jets. However, the high angular resolution of Hubble Space Telescope (HST) and Chandra X-ray Observatory (CXO) have given us the opportunity to analyze the morphological characteristics of jets at optical and X-ray photon energies, allowing for the comparison of jet morphology at different frequencies. Here we discuss the structure of large-scale quasar jets, in order to determine the origin of their X-ray radiation and the nature of the particle acceleration responsible for their multiwavelength emission.

CXO has detected about ten quasar jets with observed X-ray luminosities $L_{\rm x} \sim 10^{43} - 10^{45}$ erg s$^{-1}$ and projected lengths ranging from a few tens up to a few hundred kpc\footnote{see the web page \texttt{http://hea-www.harvard.edu/XJETS} by D. Harris; see also \citet{staw04}}. The X-ray emission is usually much too intense to be explained in terms of synchrotron radiation by the same, single-power-law electron energy distribution that produces the radio-to-optical emission. Nor can it be explained in terms of synchrotron-self Compton (SSC) radiation of a \emph{homogeneous}, few-kpc-long knot region, \emph{unless} the energy density is far above the equipartition value, which would imply extremely high jet powers \citep[e.g.,][for the case of PKS 0637-752]{cha00,sch00}. This led to the idea that the X-ray emission results from inverse-Compton up-scattering of Cosmic Microwave Background (CMB) photons --- `external inverse-Compton' (EIC) emission --- by the low-energy part of the electron energy distribution \citep{tav00,cel01,hk02}. Such a model, however, requires highly relativistic jet velocities at distances of a few hundred kpc from the active nuclei (bulk Lorentz factors $\Gamma \sim 10$), in apparent disagreement with radio observations \citep[e.g.,][]{war97}. On the other hand, the EIC model was claimed to be additionally supported by energetic considerations \citep{ghi01}. In this respect, discoveries of large-scale X-ray jets in the Gigahertz-peaked (GPS) objects PKS 1127-145 and B2 0738+313 \citep[respectively]{sie02,sie03a}, as well as in the high-redshift source GB 1508+5714 \citep[$z=4.3$,][]{sie03b} are particularly interesting. X-ray emission of large-scale quasar jets, if due to the EIC process, has also cosmological implications, as pointed out by \citet{sch02}. Therefore, it is important to discuss whether Comptonization of the CMB within highly relativistic flow is really the most plausible explanation of the CXO quasar jet observations, and to look for possibilities of discriminating between this and other models \citep[e.g.,][]{aha02,der02,sta02}. Analysis of the morphology of the emitting regions constitutes an interesting approach to this problem.

The most apparent characteristic of quasar jets is their knotty morphology with high knot-to-interknot brightness contrast, but also with distinct (in some cases) inter-knot diffuse emission. In addition, knot profiles seem to be frequency-independent, and knot extents are similar when observed at radio, optical and X-ray photon energies. Detailed observations of the 3C 273 jet \citep{jes01} reveal also that spectral changes along the flow are not correlated with brightness changes --- this may be a general characteristic of these objects. Finally, in some cases spatial offsets between the maxima of the X-ray and radio emission within the knot regions were noted \citep[e.g., PKS 1127-145,][]{sie02}. It is not clear whether all of these features can be explained in a framework of models involving extended shock waves within continuous jet flow. In fact, we argue that the morphological characteristics cannot be explained in this way, and that substantial modifications of the standard picture are required. Such modifications are especially needed if the X-ray emission of quasar jets is due to the EIC process. We propose that at least some aspects of the HST and CXO observations can be understood in terms of intermittent (modulated) jet activity. In this context, we comment on both the external-Compton scenario of quasar jet X-ray emission, and on a model of boundary layer acceleration and resulting high-energy radiation \citep{sta02}. In particular, in \S~2 we emphasize the problems with modeling X-ray knots of quasar jets as stationary regions of particle acceleration. In \S~3 we consider the possibility that the knots are moving sources of non-thermal radiation, propose a possible connection of this scenario to models of intermittent jet activity, and discuss the particle acceleration processes possibly involved in such a scenario. The discussion and final conclusions are presented in \S\S~4 and 5, respectively.

\section{Stationary knots}

In the most common version of the EIC model, in which the knots are identified with strong stationary shocks through which the jet matter flows continuously, none of the morphological features mentioned in the introduction are straightforwardly expected. The knots appear to be localized along the jet, yet stationary shocks are likely to be highly oblique (with respect to the jet axis) and significantly elongated in the flow direction, if the jet is confined by external pressure \citep{san83,kom97}. Large-scale jets in powerful radio sources display approximately cylindrical geometry, and therefore it is thought that they must be confined. The nature of their confinement is unknown, but it has been shown that the low-density and high-pressure nonthermal medium of the radio lobe can be an efficient confinement agent \citep{beg89}. If we are not seeing the shape of the shocks themselves, could the emissivity patterns give rise to the knot morphology? It seems unlikely, since the cooling time of low-energy electrons responsible for up-scattering CMB photons to the observed X-ray energies is extremely long when compared with the lifetime of electrons emitting synchrotron optical photons in an equipartition magnetic field. In addition, the EIC emission is not very sensitive to the magnetic field structure or slow adiabatic losses within the uniform, well-collimated flow. Therefore, in the case of the EIC model and a continuous jet with knot regions marking the positions of the stationary (e.g., reconfinement) shocks, one should rather expect extended X-ray knot emission, with approximately constant flux along the jet (if the jet bulk velocity is roughly constant on the large scales), and with knot profiles reflecting directly the large differences in radiative and adiabatic cooling times of electrons radiating in radio, optical and X-rays. This is in disagreement with observations, and hence modifications to the EIC model are required. One possibility is to assume an inhomogeneous structure of the emission regions, consisting of a number of small, far-from-equipartition and adiabatically expanding clumps of radiating plasma \citep{tav03}.

\subsection{EIC clumping model}

The EIC model was claimed to be strongly supported by energetic arguments \citep{ghi01}. Let us briefly re-discuss this issue in the context of the clumping scenario. If the jet is composed of a magnetic field of unknown intensity $B$ (measured in the comoving frame\footnote{Hereafter we use primes to denote quantities measured in the jet comoving frame, while unprimed quantities refer to the observer rest frame. However, we do not prime the magnetic field induction $B$ or electron Lorentz factor $\gamma$, noting instead that they always refer to the emitting plasma rest frame.}), protons, and radiating electrons, one can find a value of the magnetic field that minimizes the total kinetic power of the emission regions for a given observed synchrotron luminosity $L_{\rm syn}$, observed volume of the emitting plasma $V$, ratio of proton-to-electron energy densities $\eta \equiv u'_{\rm p}/u'_{\rm e}$, spectral energy distribution of the radiating ultrarelativistic electrons, bulk Lorentz factor $\Gamma$, and jet inclination $\theta$. This is because the total jet power is $L_{\rm tot} = L_{\rm B} + L_{\rm e} + L_{\rm p}$, where the Poynting flux is $L_{\rm B} \propto B^2$, the bulk kinetic power carried by the radiating electrons is $L_{\rm e} \propto u'_{\rm e}$, and the bulk kinetic power due to the protons is $L_{\rm p} \propto u'_{\rm p}$. As the electron energy density for a given total synchrotron luminosity is $u'_{\rm e} \propto B^{-2}$, one has also $L_{\rm e} \propto B^{-2}$ and hence $L_{\rm p} \propto \eta \, B^{-2}$. Setting
\begin{equation}
\left({\partial L_{\rm tot} \over \partial B}\right)_{B=B_{\rm cr}} = 0 ,
\end{equation}
\noindent
one can find that the magnetic field $B_{\rm cr}$ minimizing $L_{\rm tot}$ is the one for which $L_{\rm B} = L_{\rm e} + L_{\rm p}$. In other words, one gets condition for the equipartition between Ponting flux and the jet kinetic power carried by the particles (see Appendix A). It was shown \citep{ghi01} that this minimum power condition is consistent with X-ray observations of the large-scale jet associated with the quasar PKS 0637-752, if its X-ray flux is due to the EIC emission of a homogeneous knot region with radius $R \sim 10^{22}$ cm and $\Gamma \sim 10$. One should emphasize, however, that this agreement is obtained by invoking certain assumptions about the value of $\eta$ and details of the electron energy distribution. In addition, allowing for inhomogeneous structure of the emission region changes the energetic conclusions. In particular, the magnetic field that minimizes the total power for the uniform knot model, $B_{\rm cr}$, does not coincide with the minimum energy field evaluated in the framework of the clumping scenario for the radiating plasma, $\tilde{B}_{\rm cr}$. Instead, one has
\begin{equation}
\tilde{B}_{\rm cr} = f^{-1/4} \, B_{\rm cr}
\end{equation}
\noindent
where $f$ is the filling factor of the clumps (see Appendix A). Hence, if the magnetic field estimated from the observed ratio of inverse-Compton to synchrotron luminosities agrees with the critical magnetic field computed for the homogeneous knot model, $B \sim B_{\rm cr}$, then any strong clumping of the jet (i.e. $f \ll 1$) will cause large departures from the minimum power condition for the radiating plasma, $B \ll \tilde{B}_{\rm cr}$. In addition, in order to discuss energetics of the whole clumpy jet, one has to make some assumptions about the non-radiating matter between the clumps, which cannot be verified by observations. As a result, energetic arguments supporting the EIC model cannot be applied to the clumping scenario.

Losing a reliable diagnostic of jet energetics is the price one has to pay for explaining a rapid decrease of the X-ray flux outside the knots. In the expanding, radiating plasma, the magnetic field is expected to decrease in proportion to some power of the emitting region radius, $B \propto r^{-m}$, depending on the assumed magnetic field structure. Therefore, one has
\begin{equation}
{L_{\rm eic} \over L_{\rm syn}} \propto {u'_{\rm cmb} \over u'_{\rm B}} \propto r^{2m}  ,
\end{equation}
\noindent
where $L_{\rm eic}$ and $L_{\rm syn}$ are, respectively, the observed inverse-Compton and synchrotron luminosities, $u'_{\rm cmb}$ is the energy density of the CMB radiation in the comoving frame of the jet (which is constant along the flow if the bulk velocity of the expanding region is constant), and $u'_{\rm B} \propto B^2$ is the energy density of the jet magnetic field. Hence, knot profiles in the EIC model with adiabatic expansion should always be different when observed at X-ray and radio frequencies (both produced by the low-energy part of the electron spectrum), unless very rapid expansion of the radiating plasma, i.e., strong clumping of the jet, is involved. In this case, however, the contribution from synchrotron self-Compton (SSC) emission cannot be ignored. Note that the main argument in favor of the EIC emission model over the SSC model was that the latter requires clumps and departures from equipartition, in order to explain the strong X-ray emission \citep{sch00}. The second problem attributed to the EIC-clumping model was that rapid expansion of the emitting clumps would inevitably create a large population of very low energy electrons. They should in turn produce optical emission via the EIC process, possibly exceeding the synchrotron radiation (and the observational limits) at these frequencies.

Energy losses of the low-energy electrons, radiating via synchrotron radiation in the radio band and via EIC emission in the X-ray domain, are dominated by adiabatic cooling, $d \gamma / d r = - A \, \gamma / r$. This process leads to the decrease of the electron energy with linear size of the emitting region $r$ according to
\begin{equation}
\gamma \propto r^{-A}
\end{equation}

\noindent
and also to the decrease of the inverse-Compton luminosity at given observed frequency according to
\begin{equation}
L_{\rm eic} \propto r^{- A \, (s-1)} ,
\end{equation}
\noindent
where $s$ is the spectral index of the electron energy distribution (Appendix B). For $s = 2.5$ and two-dimensional expansion (i.e., $A = 2/3$), one gets a linear decrease of the observed X-ray luminosity with $r$. In such a case, in order to reduce the X-ray flux by, say, one order of magnitude, one has to allow all the emitting clumps to expand within the single knot region by, at least, one order of magnitude in their linear sizes. Thus, electrons up-scattering CMB photons to the observed $\leq$ keV energies, $\gamma \leq \delta^{-1} \, (\nu_{\rm x} / \nu_{\rm cmb})^{1/2} \sim 1000 \, \delta^{-1}$, will end up at energies $\gamma \leq 200 \, \delta^{-1}$ after such expansion, giving rise to the observed EIC emission at $\varepsilon \leq 50$ eV \citep[see][]{tav03}. This possibility, although not completely excluded, contradicts the synchrotron origin of optical emission from quasar jets, which has been confirmed repeatedly by polarization measurements. Hence, the EIC model requires a very sharp cut-off in the injected electron energy distribution and/or relatively limited expansion of the clumps --- both of which call for fine-tuning of the model parameters --- in order not to overproduce optical inverse-Compton radiation.

Small sizes of the emitting plasma in the clumpy jet can also amplify the energy density of the synchrotron photons to a level leading to domination of the SSC component over the EIC one. Note that, for the electron energy spectrum usually considered in the EIC model, the photon energy range of the SSC emission corresponds roughly with the spectral range of the EIC radiation. The ratio of observed SSC to EIC luminosity is
\begin{equation}
{L_{\rm ssc} \over L_{\rm eic}} \sim {1 \over N \, (1+z)^4} \, \left({\delta \over 10}\right)^{-6} \, \left({L_{\rm syn} \over 10^{44} \, {\rm erg/s}} \right) \, \left({R_{\rm c} \over 10 \, {\rm pc}}\right)^{-2} ,
\end{equation}
\noindent
where $N$ is the number of the emitting clumps with radius $R_{\rm c}$ per knot, $z$ is a redshift of the source and $\delta = 1 / \Gamma (1 - \beta \, \mu)$ is the jet Doppler factor for its bulk velocity $\beta = (1 - \Gamma^{-2})^{1/2}$ and inclination $\theta \equiv \cos^{-1} \mu$ (Appendix C). It is always possible to adjust the unknown parameters $\delta$, $R_{\rm c}$ and $N$ in order to obtain $L_{\rm ssc} > L_{\rm eic}$. As a result, it is possible to model the observed X-ray emission of large-scale quasar jets by SSC emission, and hence there are no obvious constraints on the large-scale jet parameters from X-ray observations. In particular, the Doppler factor of the jet in the SSC model can be smaller than that required by the EIC model.

\subsection{Synchrotron X-rays?}

Modifications to the EIC model, which are required in the case of knots treated as stationary regions of energy dissipation, decrease the model's attractiveness. Therefore, alternative scenarios postulating a synchrotron origin of the X-ray emission should be considered. However, in this case the X-ray photons would have to be produced by a different electron population than the one responsible for the radio-to-optical emission, as the X-ray flux is usually above the extrapolated radio-to-optical power-law continuum; or else deviations from a single power-law behavior in the electron energy distribution --- high energy particle pile-up in particular --- have to be invoked. Furthermore, to reconcile the observed flat X-ray spectral indices with strong cooling of the highly energetic electrons and large observed sizes of the emission regions \citep[see discussion in][]{aha02}, one must have particle acceleration occurring over extended volumes. This can be illustrated by rewriting the observed propagation length of the electrons radiating synchrotron photons with frequency $\nu_{\rm syn}$ in a form

\begin{equation}
l_{\rm syn} \sim c \, \Gamma \, t'_{\rm rad} \sim {10^{-2} \, \Gamma^{3/2} \, B_{-4}^{-3/2} \, (1+z)^{-1/2} \over 1 + 10^{-3} \, \Gamma^2 \, B_{-4}^{-2} \, (1+z)^4} \, \, \left(\nu_{\rm syn} \over 10^{18} \, {\rm Hz} \right)^{-1/2} \quad {\rm kpc} ,
\end{equation}
\noindent
where $t'_{\rm rad}$ is the radiative loss time scale, $B_{-4} \equiv B / 10^{-4}$ G, inverse-Compton radiative losses on the CMB are included and we put $\delta \sim \Gamma$ as appropriate for small jet viewing angles $\theta$ (Appendix D)\footnote{Let us mention that, as emphasized by \citet{jes01}, the apparent discrepancy between this spatial scale for the optical synchrotron emission and the observed extent of the optical jet in 3C 273 cannot be resolved solely by invoking a sub-equipartition magnetic field and/or highly relativistic velocities.}.

The turbulent second-order Fermi acceleration process provides a plausible way of obtaining high-energy particle pile-up, if the time scale for particle escape from the acceleration region is much longer than the radiative loss time scale (other possibilities will be discussed in the later sections). This is because such a process favors the particle energy for which the acceleration time scale is equal to the radiative loss time scale. With steady injection of low-energy particles, turbulent acceleration piles up radiating particles at this critical energy. As discussed by \citet{ost00} and \citet{sta02}, boundary shear layers of large-scale jets are ideal sites for such a process, as the inevitable flow instabilities very likely provide the required amount of turbulence, and a sheared magnetic field configuration does not enable radiating electrons to escape easily from the acceleration region. For conventional assumptions about the jet energetics (including energy equipartition between the magnetic field and radiating electrons), the time scales estimated by \citet{sta02} lead to sufficient pile-up of the high-energy electrons to produce synchrotron X-rays with a luminosity comparable to the observed values, without fine-tuning of the jet parameters. In addition, the continuous nature of such an acceleration process can explain the absence of cooling effects in the observed optical and X-ray continuum.

Boundary layer acceleration, as initially proposed to explain some aspects of the X-ray jet emission observed by CXO, cannot solely account for the morphological characteristics of quasar jets, and their knotty nature in particular. In this model the lifetime of the X-ray emitting electrons is extremely short, and hence the emitting region sizes are related to the extent of the particle acceleration sites and not to the spatial scales of the electron cooling, as in the EIC and SSC scenarios. Also, the interplay between localized shock acceleration and continuous, turbulent energization of the radiating electrons can result in offsets between the radio and X-ray emitting regions, while in the EIC and SSC models the maximum of the radio emission should have the same location within the knot as the maximum of the X-ray radiation. As mentioned in the introduction, spatial offsets between radio and X-ray knot maxima are observed in some large-scale quasar jets. However, a simple model of turbulent acceleration does not explain directly why the jets --- being continuous flows --- exhibit high knot-to-interknot brightness contrasts, although one can imagine some possible solutions. For example, it was suggested \citep{sta02} that the knots can form by the shock compression of the energetic electrons previously accelerated at the boundary layer upstream of the shock, in accordance with the discussion by \citet{beg90}.

\subsection{Spectral constraints}

Detailed spectral studies at optical frequencies could allow us to distinguish between an inverse-Compton and a synchrotron origin for the X-ray emission from large-scale quasar jets. For example, the presence of spectral cut-offs at infrared-to-optical photon energies would argue strongly for the EIC model, while spectral flattenings would support synchrotron models involving high-energy particle pile-ups. However, optical emission detected from a few quasar jets\footnote{see the web page \texttt{http://home.fnal.gov/\~{}jester/optjets} by S. Jester; see also \citet{staw04}}, which is most probably synchrotron in nature as indicated by polarization measurements, reveal a variety of spectral shapes, and hence no definitive statements can be made yet. In particular, optical spectral indices of the knots in powerful jets range from $\alpha_{\rm o} \sim 0.5$ in the brightest part of the 3C 273 jet \citep{jes01} up to $\geq 1.6$ in 3C 279 \citep{che02}. Interestingly, spectral hardening at ultraviolet frequencies was noted for the 3C 273 jet \citep{jes02}. These observations indicate the presence of $\gamma \sim 10^5 - 10^6$ electrons producing synchrotron infrared-to-optical photons, for an assumed equipartition magnetic field $B \sim 10^{-5}$ G, but do not give tight constraints on their spectral distribution. The broad-band energy spectrum of the radiating electrons is usually modeled in the EIC scenario as a single power-law, $n'_{\rm e}(\gamma) \propto \gamma^{-s}$, with spectral index $s \sim 2.5$ and sharp cut-offs at $\gamma_{\rm min} \sim 10^2$ and $\gamma_{\rm max} \sim 10^5 - 10^6$ \citep[e.g.,][]{sam02}. Except for its simplicity, however, such an assumption on a form of $n'_{\rm e}(\gamma)$ has neither theoretical nor observational justification. Moreover, it cannot be excluded that the electron energy distribution $n'_{\rm e}(\gamma)$ extends up to very high energies, $\gamma \sim 10^8$, as seems to be the case in a few kpc-scale FR I jets \citep[see discussion in][]{staw04}. Another point is that the broad-band radio-to-X-ray spectra for some of the knots in the large-scale jets associated with quasars 3C 273 and PKS 1136-135 cannot be explained by the EIC model, indicating with little doubt the synchrotron origin of the X-rays \citep{mar01,sam02}. Therefore, existing optical spectra of the knots in quasar jets cannot yet exclude different scenarios for modeling the radio-to-X-ray continua of these objects.

\section{Moving knots}
 
As we have discussed above, modeling knots of large-scale quasar jets as stationary regions of energy dissipation within the uniform and continuous jet flow leads to several problems in explaining the observed X-ray emission, especially in the framework of the EIC scenario. Below we explore another possibility, namely that the knots represent portions of a relativisticaly moving jet with excess kinetic power. Such a situation is expected if the activity of central engines in radio-loud AGNs is intermittent, or highly modulated. The frequency-independence of the knot sizes, as well as high knot-to-interknot brightness contrasts, can be understood if the radiating particles are present mainly within the separate portions of the jet flow --- the knots --- as compared with the interknot regions. This is true almost independently of the emission process involved, and in particular it applies also to the EIC model. In addition, in moving knot scenarios the brightness changes along the jet path do not have to correlate with the spectral changes, if they are controlled by distinct physical processes, i.e., if the former are due to changes in the kinetic energy of the flow resulting from intermittent or modulated jet activity, while the latter are determined by particle acceleration and cooling processes. These facts make the idea of intermittent jet activity promising for explaining the morphologies of CXO quasar jets.

\subsection{Modulated jet activity}

The idea of intermittent jet activity was presented in the context of `partial' radio jets associated with the powerful radio galaxies 3C 219 \citep{bri86,cla92} and 3C 288 \citep{bri89}. It was also proposed that the restarting jet model can, in a natural way, reconcile the presence of the extended radio emission observed in a few GPS sources (e.g., 0108+388) with the inferred young ages of GPS compact radio structures, $< 10^4$ yrs \citep{bau90,sta90}. We note that extended radio emission around GPS objects can sometimes reach Mpc scales, as in the cases of B1144+352 \citep{sch99} and B1245+676 \citep{mar03a}. In addition, as discussed by \citet{rey97}, intermittent jet activity can explain the general overabundance of compact radio objects when compared with the number of classical (extended) doubles. In particular, \citet{rey97} showed that the observed distribution of sizes for extragalactic radio sources can be explained in the framework of a simple evolution scenario if one assumes that the jet-like active period persists only over $\sim 10^4$ years, and refreshes on the time scale of $\geq 10^5$ years. Interestingly, similar time scales can be obtained in the thermal-viscous instability model for the accretion disk \citep{sie97,jan03}. This suggests that the jet activity --- and hence the whole evolution of the radio source --- could be driven by unstable accretion disks surrounding supermassive black holes \citep[see in this context][]{mar03b}.

Non-steady jet activity, albeit on longer time scales than discussed above, was also suggested in order to explain some morphological features observed in the so-called Giant Radio Galaxies \citep[linear sizes of the order of Mpc;][]{sub96}. In a few sources belonging to this class, intermittent jet activity was confirmed directly by observations of double-double radio structure \citep{sch00a}. In particular, observations of B1834+620 \citep{sch00b}, J0116-473 \citep{sar02}, and B1545-321 \citep{sar03} show a pair of `inner' radio jets which propagate within a diffuse radio-emitting medium formed by previous jet activity, and which display FR II radio morphology. Such inner radio structures are characterized by lower radio luminosity, larger axis ratios of the lobes, and weaker hot-spots than classical doubles with similar sizes, in agreement with theoretical predictions for restarting bipolar jets \citep{cla91,kai00}. 

A restarting jet propagates not within the unperturbed intergalactic or intracluster medium of the active galaxy, but within the interior of the radio lobe created by the previous jet activity \citep[e.g.,][]{beg89}. The number of double-double radio sources is small, probably because this kind of morphology can be formed only if the new pair of jets is created not earlier than $10^7 - 10^8$ yrs after the old ones, in order to allow for an increase in the density of the cold gas in the relict radio cocoon \citep{kai00}. Otherwise, the rarefied and mostly nonthermal medium formed during the previous period of jet activity prevents formation of new lobes and hot-spots\footnote{On the other hand, the time scale of the quiescent periods in double-double sources cannot be longer than $10^8$ yrs, because after this time the old lobe structure is not expected to remain visible \citep{kom94}.}. For the quasar jets considered here, we require the time scale between the two subsequent jet active epochs to be not longer than $10^5$ yrs. This is because the linear sizes of the knot regions are consistent with a $10^4$-year duration of jet activity, while distances between subsequent knots are on average consistent with quiescent periods an order of magnitude longer. Therefore, we characterize the quasar jets in terms of modulation of the jet kinetic power, rather than in terms of restarting flows. This idea is strongly supported by radio observations of such structures \citep[see, e.g.,][]{bri94}. The respective time scales for activity and quiescence can differ from object to object, but on average they seem to be in agreement with the ones postulated above. In some sources, however, they can be smaller than average, resulting in a more continuous flow (as in 3C 273?).

We also note that Galactic X-ray transients are known to produce radio emission which is widely modeled in terms of multiple jet-like ejection events \citep[see review by][]{fen03}. Modulated jet activity of the large-scale jets considered here constitutes an interesting link between quasar and microquasar phenomena, although the physical processes controlling intermittent/highly modulated activity of the central engines in both types of sources can be different. As a result, the time scales for the active and quiescent periods in galactic and extragalactic jet sources do not have to scale in a simple way with the mass of the central compact object.

\subsection{Energy dissipation and particle acceleration}

The moving portions of the jet have to be sites of particle acceleration. \citet{tav03} rejected this possibility, arguing that either continuous or episodic particle acceleration acting within moving knots cannot reproduce the required electron energy distribution. However, as the synchrotron spectrum at frequencies higher than the radio band is unknown in most large-scale quasar jets, such a statement is premature (see \S~2.3). Note that \citet{tav03} base their conclusions on the \emph{assumed} electron energy distribution. In reality, many different processes can control the spectral behavior of radiating electrons, and it is not certain --- from both observational and theoretical points of view --- how well a single power-law describes the spectrum. In the framework of the intermittent jet activity model considered in this section, turbulent particle acceleration processes due to interaction of the jet with the surrounding medium are expected to lead to the formation of a piled-up distribution of energetic electrons radiating synchrotron X-rays. The formation of internal shocks within the relativistic, highly modulated jet flow is also very probable. In the latter case, one can expect the EIC process to dominate production of the X-rays, whereas the deominant radiation mechanism is less clear in the former. Realistically, the situation can be quite complex, involving an ensemble of multiple shocks (formed due to both velocity irregularities and propagation instabilities of the flow) as well as the interplay between coherent and stochastic particle acceleration processes.

\subsubsection{Internal shocks}

The formation of shocks in extragalactic jets is most likely related to intrinsic velocity irregularities of the flow, caused by the non-steady operation of the central engine, as proposed by \citet{ree78}. Such internal shocks are not stationary, but instead are moving together with the radiating matter, possibly with highly relativistic bulk velocities. Internal shocks are typically characterized by a low efficiency of converting the beam kinetic energy to internal energy of radiating particles \citep[see, e.g.,][]{laz99,spa01}. This is a very convenient situation for the objects considered here, because their observed radiative luminosities, if corrected for Doppler beaming effects, are several orders of magnitude smaller than the estimated jet kinetic powers ($L_{\rm j} \sim 10^{47}$ erg s$^{-1}$), and hence the fraction of the kinetic energy flux of the jet dissipated to the radiating particles can be small.

Let us consider the simplest situation, in which two colliding portions of the jet flow differ only in bulk velocity. Then, for respective bulk Lorentz factors $\Gamma_2 > \Gamma_1 \gg 1$, a double shock structure (symmetric in the comoving frame of the contact discontinuity for the same densities $\rho_1 = \rho_2$) develops. The bulk Lorentz factor of the contact discontinuity, which is also the Lorentz factor of the shocked radiating plasma, is then
\begin{equation}
\Gamma_{\rm ct} \sim \sqrt{\Gamma_1 \, \Gamma_2} .
\end{equation}
\noindent
Knowing solely $\Gamma_1$ and $\Gamma_2$ (and assuming an equation of state for the downstream matter), one can estimate velocities of the forward and reverse shocks from standard shock-jump conditions \citep[Appendix E]{bla76}. The extent of the shocked region along the jet, in the contact comoving frame denoted by primes, is
\begin{equation}
\Delta l' \sim 2 \, c \, |\beta'_{\rm sh}| \, \Delta t' ,
\end{equation}
\noindent
where $|\beta'_{\rm sh}|$ is the velocity of the forward and reverse shock fronts and $\Delta t'$ is the time since formation of the double-shock structure. Observations of the knots at the maximum synchrotron frequency $\nu_{\rm max}$ can give some constraints on the maximum value for $\Delta l'$. This is because the time interval $\Delta t'$ cannot be larger than the radiative loss time scale for the most energetic electrons radiating at $\nu_{\rm max}$, in order to ensure activity of the whole shocked region at this frequency. Therefore, for a small viewing angle of the jet, the observed extent of the shocked region can be limited according to
\begin{equation}
\Delta l \leq l_{\rm max} \equiv 2 \, c \, |\beta'_{\rm sh}| \, \Gamma_{\rm ct} \, t'_{\rm rad}(\nu_{\rm max}) ,
\end{equation}
\noindent
where $t'_{\rm rad} = t'_{\rm rad}(\nu_{\rm max})$ is the radiative loss time scale for a given synchrotron frequency $\nu_{\rm max}$ (Appendix D). Figures 1--3 show the parameter $l_{\rm max}$ as a function of the magnetic field strength in the radiating plasma, for $\nu_{\rm max} = 10^{15}$ Hz and different bulk Lorentz factors $\Gamma_1$ and $\Gamma_2$. Note that, for a wide range of bulk Lorentz factors of the colliding portions of the jet, the extent of the double-shock structure in the observer frame cannot be larger than $\sim 1$ kpc, independent of the magnetic field intensity. This value, interestingly, is consistent with the observed lengths of the optical knots in quasar jets. Therefore, if dynamical time scales controlling the development of the double-shock structure ensure that $\Delta l \leq l_{\rm max}$ for the maximum synchrotron frequency occurring in the optical band, then the EIC/internal shock scenario can explain frequency-independent knot profiles.

However, there are two important problems with the model discussed above. The first is that radio polarization of the quasar jets indicates a magnetic field which is on average parallel to the jet axis. As an example we mention the jet in PKS 0637-752, where the X-ray emission as observed by CXO is strong only in the regions characterized by a longitudinal magnetic field \citep{sch00}. Such a situation is inconsistent with the internal shock model, because compression of the magnetic field should result in a transverse orientation (Laing 1980). The second problem is the fine-tuning required between the dynamical time scales controlling development of the double-shock structure, and the radiative loss time scales of the radiating particles, in order to obtain almost universal extents of the optical/X-ray knots, $\sim 1$ kpc.

\subsubsection{Multiple oblique shocks and turbulent acceleration}

We now discuss particle acceleration processes related to ensembles of shocks which are not comoving with the radiating matter, i.e., have a different pattern velocity than the jet. Numerical simulations reveal the presence of such a structure at the jet boundary \citep[e.g.,][]{nor82}, although it is possible for these shocks to encompass the whole jet width.  The particle acceleration process related to the ensemble of many shocks can be different from single shock acceleration. This issue was considered in the literature in different contexts starting from \citet{bla80}, and discussed further by \citet{spr88,ach90,sch93,mel93,pop94,mel97,mar99} and \citet{gie00}. Physically, what distinguishes between single and multiple shock acceleration is that in the latter case, particles can change their momenta between subsequent shock events due to adiabatic and radiative losses, and can also escape from the regions between of the shocks, thus avoiding further acceleration. As shown by \citet{sch93} and \citet{mar99}, the interplay between time scales representing radiative cooling, particle escape and advection can result in the formation of different particle energy spectra, with spectral flattenings, curvatures, cut-offs or high-energy bumps (spectral pile-ups).

Let us consider the appropriate time scales for multiple shock acceleration in the context of the knots in quasar jets. Following \citet{sch93}, we assume that the time scale for acceleration and subsequent decompression at each shock is much shorter than the time scales connected with radiative losses, $t'_{\rm rad}$, and particle escape from the region between subsequent shocks (i.e., the region where the nonthermal radiation is produced), $t'_{\rm esc}$, as well as the interval between the subsequent shock acceleration events, $t'_{\rm b}$. Then, the average particle energy distribution resulting from multiple (roughly identical) shock acceleration events depends on the ratios of these three characteristic time scales. Note first, that if particle escape is controlled by diffusion from the region between the shocks, which is characterized by a thickness $D$, then the condition $t'_{\rm esc} > t'_{\rm rad}$ reads 
\begin{equation}
D > 0.04 \, B_{-4}^{-3/2} \quad {\rm pc}
\end{equation}
\noindent
for strong turbulence, where the particle mean free path is comparable to the particle gyroradius, and for inverse-Compton losses comparable to synchrotron losses (Appendix F). Thus, if $D$ is of order of the jet radius ($\sim 1$ kpc), then particle escape is inefficient compared to the radiative losses, independent of particle energy. Also, if the mean separation of the shocks is less than $\sim 1$ kpc, particles cannot escape before passing through many shocks, and therefore they experience multiple (and not single) shock acceleration. In such a case, as shown by the analysis of \citet{sch93}, one should expect the formation of a flat power-law spectrum for particles satisfying $t'_{\rm rad} > t'_{\rm b}$, a steep power-law spectrum (but not necessarily an exponential cut-off) for particles satisfying $t'_{\rm rad} < t'_{\rm b}$, and a spectral pile-up at particle energies for which $t'_{\rm rad} \sim t'_{\rm b}$.

Now consider what the shock separation has to be in order to obtain a piled-up distribution of particles emitting the observed synchrotron keV photons. If the radiating plasma between the shocks is highly relativistic (with the respective bulk Lorentz factor $\Gamma \gg 1$), then the condition $t'_{\rm rad} \sim t'_{\rm b}$ reads
\begin{equation}
d' \sim c \, t'_{\rm rad} ,
\end{equation}
\noindent
where $d'$ is the shock separation as measured in the radiating plasma rest frame (Appendices D and F). The parameter $d'$ as a function of $\Gamma$, for the observed critical synchrotron photon energy $h \nu_{\rm cr} \sim 1$ keV and different magnetic field intensities, is shown on figure 4. For $\Gamma \sim 3 - 10$ and $B \sim 10^{-5} - 10^{-4}$ G one gets $d' \sim 10 - 100$ pc. For larger shock separations, pile-ups are expected to occur at lower energies of the synchrotron photons. However, these values should be considered as illustrative only, because the analysis above refers to a very simplified situation, in which we neglect the effects of turbulent second-order Fermi acceleration and assume identical properties for all the shocks. For reasonable large-scale jet parameters, however, an ensemble of shock waves present within the extended knot region can result in the formation of a flat power-law electron energy distribution followed by a high-energy pile-up announced by its X-ray synchrotron emission. This situation is similar to the model discussed by \citet{sta02}, but here the time scale between subsequent shock events, and not the second-order Fermi acceleration time scale, $t'_{\rm F \, II}$, determines the critical electron energy. The ratio of these two time scales, for $B_{-4} \sim 1$ and electron Lorentz factor $\gamma \sim 10^8$, can be rewritten as
\begin{equation}
{t'_{\rm b} \over t'_{\rm F \, II}} \sim \left({\beta_{\rm A} \over 0.01} \right)^2 \, \left( {d' \over 10 \, {\rm pc}} \right)
\end{equation}
\noindent
where $V_{\rm A} \equiv \beta_{\rm A} c$ is the Alfven speed in the radiating plasma (Appendix F). Therefore, we conclude that an ensemble of oblique shock waves within the knot regions of large-scale quasar jets can act in the analogous way to second-order Fermi acceleration, as discussed by \citet{sta02}. Both processes, which in a realistic situation can be hard to separate, are due to the interaction of the propagating jet with the surrounding medium. Both of them can result (for reasonable jet parameters) in a pile-up of high-energy electrons radiating X-ray synchrotron photons, and both are consistent with the observed longitudinal magnetic field.

\section{Discussion}

Inverse-Compton scattering of the CMB radiation field within a highly relativistic flow is commonly believed to explain relatively strong X-ray emission detected by CXO from large-scale quasar jets. This model, as initially proposed, is attractive because of its simplicity and energetic efficiency. On the other hand, it is inconsistent with morphological properties of these objects, namely with the frequency-independent knot profiles, \emph{if} the jet flow is continuous and the knots are stationary (e.g., if they are manifestations of the reconfinement shocks through which the jet matter is flowing at a constant rate). It was proposed \citep{tav03} that the inconsistency can be resolved by assuming that the radiating jet region consists of small, adiabaticaly expanding clumps. However, as discussed in \S~2.1, by invoking an inhomogeneous structure of the emitting regions, one loses the advantages of the EIC model over other models proposed in the literature. In particular, when clumps are incorporated into the EIC model, emitting regions generally have to be far from the minimum power condition. In this situation, it becomes much more difficult to arrange for the radiating particles to be confined to the required volumes. Also, knot profiles will be observed as almost frequency-independent only if the radiating regions expand very rapidly, starting from very compact clumps. Such a situation leads to the problem that contributions from the SSC process cannot be excluded anymore. Furthermore, fine-tuning of the parameters is required in order not to overproduce optical emission by the EIC process at the final stage of the clump evolution. One should emphasize that, if the X-rays from large-scale quasar jets --- modeled as continuous flows with localized and stationary particle acceleration sites --- are indeed inverse-Compton in origin, then clumping of the knot regions is unavoidable. In the case of the EIC scenario this is because strong adiabatic losses are required in order to obtain knots instead of uniform and continuous emission along the jet, while in the case of the SSC process it is because the energy density of synchrotron photons within the emitting regions is required to be larger than in the homogeneous case in order to provide the observed level of X-ray emission. Therefore, we consider the clumpy jet model (as proposed by \citet{tav03}) as unlikely.
 
We note that complex substructure of the knots is suggested by the time variability of the kpc-scale jet in the FR I radio galaxy M 87 \citep[e.g.,][]{har97,har03}. In FR I sources, the X-ray emission observed by CXO is most probably synchrotron in origin. Also, detailed studies of the Centaurus A jet radio and X-ray radiation seem to indicate a spine-boundary layer morphology \citep{hrd03}. This can be taken as an indication that dissipation processes acting at the jet boundary can be crucial for understanding the high-energy radiative output of extragalactic large-scale jets, and their X-ray emission in particular \citep[ \S~2.2]{sta02}. We stress that the nature of the quasar jet clumps --- as localized small sites of particle acceleration --- is unknown. It was proposed \citep{tav03} that the clumpy structure can form when the jet crosses a region filled with small clouds of ambient medium, or that the clumps can be related to magnetic reconnection events.

Regardless of the specific mechanism of X-ray production, the morphological properties of quasar jets (inferred from multiwavelength observations) can be understood if the knots are not stationary regions of energy dissipation within the continuous jet flow, but instead are moving, separate portions of the jet matter. In \S~3.1 we outlined the possible connection between this idea and highly modulated jet activity. In this case, a variety of particle acceleration processes (including reconnection, acting in a way analogous to second-order Fermi acceleration) can play a role. Unfortunately, optical observations, which could be crucial in discriminating among models, cannot give definitive answers at the moment. Therefore, detailed spectral and morphological studies are still needed. Let us mention in this context one especially interesting object, which could be a very promising target for such multiwavelength studies: the large-scale jet associated with the Seyfert galaxy 3C 120. In this object, intermittent jet activity seems to take place, while the X-ray emission of the knot seems to be exceptionally inconsistent with the EIC model. 

3C 120 ($z = 0.033$) displays large-scale radio structure, with a one-sided bending jet (observed up to roughly $100$ kpc from the core) and diffuse, amorphous radio emission extending in several directions up to $> 400$ kpc \citep{wal87}. VLBI observations reveal complex jet structure with superluminal motions of some jet components at the $1 - 100$ pc scale \citep[e.g.,][]{wal01}. This, together with the large bending of the flow at kpc-scales, indicate a small jet inclination to the line of sight. At $\sim 2 - 3$ kpc from the host galaxy, an elongated radio knot, oriented perpendicular to the flow axis, is observed. Long-term VLA monitoring \citep{wal97} indicates that this knot is stationary (or moving only subluminally). Farther away the jet bends and fades, but is manifested again at $\sim 20''$ from the core as a bright, extended radio knot, oriented obliquely with respect to the jet path. The radio spectral index $\alpha_{\rm r} \sim 0.65$ remains constant along the flow, although brightness variations are large. Optical observations reveal polarized emission aligned with some parts of the radio jet in the first few kpc from the core, but with morphology not corresponding directly to the radio one \citep{hjo95}. Surprisingly, the $20''$ knot turn out to possess an X-ray counterpart with a $0.2 - 2$ keV luminosity $\sim 10^{42}$ erg s$^{-1}$, as shown by ROSAT observations \citep{har99}. Optical upper limits (placed below the interpolated radio-to-X-ray continuum) allow one to estimate the optical-to-X-ray power-law slope for this region as $\alpha_{\rm o-x} < 0.35$.

The EIC model applied to the X-ray knot of the 3C 120 jet gives an uncomfortably small viewing angle of the jet, $\theta \sim 1.5^0$, and a very high bulk Lorentz factor $\Gamma \sim 40$ \citep{hk02}. Note that, in such a case, the deprojected distance of the X-ray knot from the host galaxy is on the order of Mpc. On the other hand, radio observations together with optical upper limits exclude the possibility that the radio-to-X-ray continuum results from the synchrotron emission of a single power-law electron energy distribution. Also, if the X-ray emission is synchrotron radiation, for example due to an additional electron population as suggested by \citet{har99}, the flat spectral index and large extent of the knot call for continuous particle energization\footnote{\citet{aha02} proposed an alternative explanation, involving synchrotron emission of very high energy protons.}. We suggest that the X-ray emission of the knot in the 3C 120 jet is indeed synchrotron in origin, and is a manifestation of the pile-up of high-energy electrons by some particle acceleration process acting continuously within the radiating region. This region can be further identified with a separate portion of the jet matter, possibly moving with relativistic velocities (but not as high as required by the EIC model), ejected from the active nucleus some $10^5$ yrs ago. This is consistent with the morphology of the 3C 120 extended radio lobes, which also indicate intermittent jet activity.

\section{Conclusions}

We propose that the knots observed in large-scale quasar jets in radio, optical and X-rays represent portions of the jet with an excess of kinetic power due to intermittent high activity of the jet engine. The appropriate time scales for jet activity and quiescent epochs ($\sim 10^4$ yr and $\sim 10^5$ yr, respectively) are roughly consistent with the ones estimated in different contexts for powerful radio sources \citep{rey97,sie97}. The proposed scenario can explain some morphological properties of quasar jets, independently of the exact emission process responsible for production of X-rays. Indeed, the dominant emission process cannot be designated for certain yet. It is possible that the X-ray emission of quasar jets is due to inverse-Compton scattering of the CMB by low-energy electrons accelerated in internal shocks. On the other hand, it is also possible that the observed keV photons are synchrotron in origin, being a manifestation of high-energy electrons piled-up at the critical energy $\sim 100$ TeV due to the second-order Fermi process and/or multiple shock acceleration. The fact that an exceptionally high Doppler factor is required to explain the 3C 120 X-ray data in the framework of the EIC scenario, suggests that synchrotron emission is likely to be the dominant process in this object, at least. Whatever emission mechanism prevails in most large-scale quasar jets, intermittent jet activity assures, in a natural way, high knot-to-interknot brightness contrasts, frequency-independent knot profiles, and almost universal extents of the knots ($\sim 1$ kpc). 
\acknowledgments

\L S, MS and MO were supported by the grant PBZ-KBN-054/P03/2001. \L S acknowledges also financial support from Fundacja Astronomii Polskiej. MCB acknowledges support from National Science Foundation grant AST-0307502. A part of this work was done during the stay of \L S and MS at JILA, University of Colorado. 

\appendix
\section{Minimum power condition}

The total power of a cylindrical relativistic jet with bulk Lorentz factor $\Gamma > 1$ and radius $R$, consisting of magnetic field with comoving energy density $u'_{\rm B} \equiv B^2 / 8 \, \pi$, radiating ultrarelativistic electrons with the comoving energy density $u'_{\rm e}$ and protons with $u'_{\rm p} \equiv \eta \, u'_{\rm e}$, is
\begin{equation}
L_{\rm tot} = L_{\rm B} + L_{\rm e} + L_{\rm p}
\end{equation}
\noindent
where the Poynting flux is
\begin{equation}
L_{\rm B} = \pi R^2 \, c \Gamma^2 \, u'_{\rm B}
\end{equation}
\noindent
and the power carried by electrons and protons is, respectively
\begin{equation}
L_{\rm e} = \pi R^2 \, c \Gamma^2 \, u'_{\rm e} \quad {\rm and} \quad L_{\rm p} = \eta \, L_{\rm e} .
\end{equation}
\noindent
For a given electron energy distribution (i.e., the respective moments $\langle \gamma \rangle$ and $\langle \gamma^2 \rangle$), the electron energy density can be related to synchrotron luminosity of some radiating volume $V'$ by
\begin{equation}
L'_{\rm syn} = {4 \over 3} \, {c \sigma_{\rm T} \over m_{\rm e} c^2} \, {\langle \gamma^2 \rangle \over \langle \gamma \rangle} \, u'_{\rm B} \, V' \, u'_{\rm e} .
\end{equation}
\noindent
Hence
\begin{equation}
L_{\rm e} = 6 \pi^2 \, {m_{\rm e} c^2 \over \sigma_{\rm T}} \, {\langle \gamma \rangle \over \langle \gamma^2 \rangle} \, {R^2 \, \Gamma^2 \over B^2} \, {L'_{\rm syn} \over V'} .
\end{equation}
\noindent
For a radiating region modeled either as a moving or steady source, one has the transformation to the observer frame
\begin{equation}
{L'_{\rm syn} \over V'} = {1 \over \delta^3} \, {L_{\rm syn} \over V}
\end{equation}
\noindent
where the Doppler factor of the jet emission is $\delta \equiv 1 / \Gamma (1 - \beta \mu)$ and $\theta \equiv \cos^{-1} \mu$ is the jet viewing angle \citep[see, e.g.,][]{sik97}. Note that, for small jet viewing angles ($\delta \sim \Gamma$), the Poynting flux and the bulk power of radiating electrons do not behave in opposite ways with respect to $\Gamma$ and $B$, because $L_{\rm B} \propto \Gamma^2 \, B^2$ while $L_{\rm e} \propto \Gamma^{-1} \, B^{-2}$ \citep[cf.][]{ghi01}. Setting 
\begin{equation}
\left({\partial L_{\rm tot} \over \partial B}\right)_{B=B_{\rm cr}} = 0
\end{equation}
\noindent
one can find that the magnetic field $B_{\rm cr}$ minimizing $L_{\rm tot}$ is the one for which $L_{\rm B} = L_{\rm e} + L_{\rm p}$. This value is given by
\begin{equation}
B_{\rm cr}^4 = 36 \pi^2 \, {m_{\rm e} c^2 \over c \, \sigma_{\rm T}} \, {\langle \gamma \rangle \over \langle \gamma^2 \rangle} \, {L_{\rm syn} \over \delta^3 \, V} \, (1 + \eta) .
\end{equation}
\noindent

Let us next consider a clumpy jet, consisting of $N$ identical clumps of radiating matter with comoving volumes $V'_{\rm c}$, which are present within the region of volume $V'$. The clump filling factor can be defined as
\begin{equation}
f \equiv {N \, V'_{\rm c} \over V'} .
\end{equation}
\noindent
Assuming that the jet velocity is uniform within the jet region under consideration, the Poynting flux for the radiating matter is then
\begin{equation}
\tilde{L}_{\rm B} = \pi R^2 \, c \Gamma^2 \, f \, \tilde{u}'_{\rm B} .
\end{equation}
\noindent
This follows from the fact that, given the magnetic energy of each clump $E'_{\rm B, \, c} \equiv \tilde{u}'_{\rm B} \, V'_{\rm c}$, the total magnetic energy of the radiating matter flowing through the jet surface $\pi R^2$ is $N \, E'_{\rm B, \, c} / V' = f \, \tilde{u}'_{\rm B}$. Similarly, the bulk kinetic powers of the emitting electrons and protons (for an assumed $\eta$ the same as in the homogeneous knot model) are, respectively,
\begin{equation}
\tilde{L}_{\rm e} = \pi R^2 \, c \Gamma^2 \, f \, \tilde{u}'_{\rm e} \quad {\rm and} \quad \tilde{L}_{\rm p} = \eta \, \tilde{L}_{\rm e} .
\end{equation}
\noindent
However, one has also
\begin{equation}
\tilde{u}'_{\rm e} \propto {L'_{\rm syn, \, c} \over V'_{\rm c}} = {L'_{\rm syn} \over f \, V'}
\end{equation}
where $L'_{\rm syn, \, c} = L'_{\rm syn} / N$ is the intrinsic synchrotron luminosity of each clump. As a result, the bulk kinetic energy of the particles computed for the radiating regions of the clumpy jet is the same as that computed for the homogeneous knot model. Hence, the magnetic field minimizing the total power of the radiating clumpy matter is
\begin{equation}
\tilde{B}_{\rm cr} = f^{-1/4} \, B_{\rm cr} .
\end{equation}
\noindent

\section{Adiabatic losses in the EIC model}

Adiabatic losses of ultrarelativistic electrons with Lorentz factor $\gamma$, within an expanding region of linear size $r$, obey 
\begin{equation}
{d \gamma \over d r} = - A \, {\gamma \over r}
\end{equation}
\noindent
(where $A = 1$ [2/3] for three-dimensional [two-dimensional] expansion), and lead to the evolution of the electron distribution according to
\begin{equation}
N'(\gamma, \, r) = \left( {r \over r_0} \right)^{A} \, N'_0(\gamma_0) .
\end{equation}
\noindent
Here $N'(\gamma, \, r) \equiv n'_{\rm e}(\gamma) \, V' \,$, $n'_{\rm e}(\gamma) \propto \gamma^{-s}$ is the electron energy distribution, $N'_0(\gamma_0)$ is the number of electrons injected impulsively at $r = r_0$ with initial energies $\gamma_0$, and
\begin{equation}
\gamma = \left({r \over r_{0}}\right)^{-A} \, \gamma_0
\end{equation}
\noindent
\citep[see, e.g.,][]{ato99}.

The inverse-Compton luminosity due to electrons with given energy $\gamma_{\rm eic}$ is
\begin{equation}
[\nu L_{\nu}(r)]_{\rm eic} \propto n'_{\rm e}(\gamma_{\rm eic}) \, V' = N'(\gamma_{\rm eic}, \, r ) . 
\end{equation}
\noindent 
Therefore, the adiabatic decrease of the EIC luminosity at a given frequency $\nu_{\rm eic} \propto \gamma_{\rm eic}^2$ is
\begin{equation}
{[\nu L_{\nu}(r)]_{\rm eic} \over [\nu L_{\nu}(r_0)]_{\rm eic}} = {N'(\gamma_{\rm eic}, \, r) \over N'_0(\gamma_{\rm eic})} = \left({r \over r_0}\right)^{-A \, (s-1)}   . 
\end{equation}
\noindent

\section{SSC emission in the clumping model}

The ratio of the observed SSC luminosity to the observed EIC luminosity is
\begin{equation}
{L_{\rm ssc} \over L_{\rm eic}} = {4 \over 3} \, \left({1+\beta \over 1+\mu}\right)^2 \, {\Gamma^2 \over \delta^2} \, {\tilde{u}'_{\rm syn} \over u'_{\rm cmb}}
\end{equation}
\noindent
\citep[e.g.,][]{sta03}. The comoving energy density of the CMB radiation at redshift $z$ is
\begin{equation}
u'_{\rm cmb} = {4 \over 3} \, \Gamma^2 \, a T^4 \, (1+z)^4
\end{equation}
\noindent
where $a T^4 \sim 4 \times 10^{-13}$ erg cm$^{-3}$. The comoving energy density of the synchrotron emission in the clumpy jet, for clumps with radius $R_{\rm c}$, is
\begin{equation}
\tilde{u}'_{\rm syn} = {L'_{\rm syn, \, c} \over 4 \pi \, R_{\rm c}^2 \, c} .
\end{equation}
\noindent
The intrinsic synchrotron luminosity of each clump, $L'_{\rm syn, \, c}$, is related to the observed synchrotron luminosity of the whole knot region, $L_{\rm syn}$, by
\begin{equation}
L'_{\rm syn, \, c} = {L_{\rm syn} \over N \, \delta^4}
\end{equation}
where $N$ is a number of identical clumps. Hence, one finally obtains
\begin{equation}
{L_{\rm ssc} \over L_{\rm eic}} = {(1+z)^{-4} \over 4 \pi \, c \, a T^4} \, \left({1+\beta \over 1+\mu}\right)^2 \, {L_{\rm syn} \over \delta^6 \, N \, R_{\rm c}^2} .
\end{equation}
\noindent

\section{Radiative energy losses}

The time scale of radiative losses via synchrotron and inverse-Compton processes, where the latter is dominated by Comptonization of the CMB, as measured in the comoving frame of the radiating plasma is
\begin{equation}
t'_{\rm rad} \sim { 3 \, m_{\rm e} c^2 \over 4 \, c \sigma_{\rm T}} \, { \gamma^{-1} \over u'_{\rm B} \, (1 + X)}
\end{equation}
\noindent
where $X \equiv u'_{\rm cmb} / u'_{\rm B}$ (see Appendix C). The electron Lorentz factor $\gamma$ is related to the observed frequency of the synchrotron emission, $\nu$, by
\begin{equation}
\nu = 0.29 \cdot {3 \, e \over 4 \pi \, m_{\rm e} c} \, {\delta \, B \over (1+z)} \, \gamma^2
\end{equation}
\noindent
and therefore, for a given observed synchrotron frequency $\nu_{\rm syn}$, one obtains
\begin{equation}
t'_{\rm rad} \sim {\sqrt{8 \pi \, m_{\rm e} c \, e} \over \sigma_{\rm T}} \, \, {\delta^{1/2} \, B^{-3/2} \, (1+z)^{-1/2} \over 1 + {32 \pi \over 3} \, \Gamma^2 \, a T^4 \, B^{-2} \, (1+z)^4} \, \, \nu_{\rm syn}^{-1/2} . 
\end{equation}
\noindent
Due to inverse-Compton losses, for a given $\nu_{\rm syn}$ the propagation length of the synchrotron-emitting electrons as measured in the observer rest frame, $l_{\rm rad} \sim c \, \Gamma \, t'_{\rm rad}$, cannot be arbitrary large, and passes through a maximum as a function of $\Gamma $ or $B$ \citep{jes01}.

\section{Internal shocks}

If the colliding portions of the jet flow differ only in bulk velocities ($\beta_1$ and $\beta_2$, respectively), one can find the bulk velocity of the contact discontinuity from the energy and momentum conservation laws:
\begin{equation}
\beta_{\rm ct} = {\beta_1 \, \Gamma_1 + \beta_2 \, \Gamma_2 \over \Gamma_1 + \Gamma_2}
\end{equation}
\noindent
\citep[e.g.,][]{kom97}. Hence, for highly relativistic velocities one has the contact discontinuity bulk Lorentz factor
\begin{equation}
\Gamma_{\rm ct} \sim \sqrt{\Gamma_1 \, \Gamma_2} . 
\end{equation}
\noindent
After the collision, a double-shock structure develops. This structure is symmetric in the comoving frame of the contact discontinuity (denoted by primes), and therefore the extent of the shocked region along the jet is
\begin{equation}
\Delta l' = 2 \, c \, |\beta'_{\rm sh}| \, \Delta t'
\end{equation}
\noindent
where $|\beta'_{\rm sh}|$ is the shock velocity in the contact frame and $\Delta t'$ is the time since the collision. In order to find the shock velocity, one has to solve the shock-jump conditions between the upstream plasma, moving with velocity $\beta_1$, and the downstream plasma, moving with velocity $\beta_{\rm ct}$. This can be done conveniently in the comoving frame of the upstream plasma (denoted by $(1)$), to obtain the Lorentz factor of the shock
\begin{equation}
\Gamma_{\rm sh}^{(1)} = \sqrt{{(\Gamma_{\rm ct}^{(1)}+1) \, [\hat{\gamma}_{\rm ct} \, (\Gamma_{\rm ct}^{(1)}-1) + 1]^2 \over \hat{\gamma}_{\rm ct} \, (2 - \hat{\gamma}_{\rm ct}) \, (\Gamma_{\rm ct}^{(1)} - 1) + 2}}
\end{equation}
\noindent
\citep{bla76}. The constant $\hat{\gamma}_{\rm ct}$ can be identified with the ratio of specific heats of the upstream matter (we assume $\hat{\gamma}_{\rm ct} = 5/3$). Next, the shock velocity $\beta'_{\rm sh}$ in the contact frame can be found by transforming $\Gamma_{\rm sh}^{(1)}$ to the downstream plasma frame. For large $\Gamma_1$ and $\Gamma_2$, $|\beta'_{\rm sh}|$ is expected to be subrelativistic.

The shocked region is a moving radiating portion of the jet flow, and therefore its deprojected observed length is
\begin{equation}
\Delta l = \delta_{\rm ct} \, \Delta l'
\end{equation}
\noindent
With the requirement $\Delta t' \leq t'_{\rm rad}$, where $t'_{\rm rad}$ is the lifetime of the radiatively cooling electrons in the downstream plasma moving with bulk Lorentz factor $\Gamma_{\rm ct}$ (Appendix D), one can write
\begin{equation}
{\Delta l \over \delta_{\rm ct}} \leq 2 \, c \, |\beta'_{\rm sh}| \, t'_{\rm rad}
\end{equation}
\noindent
to obtain for small jet inclinations (i.e., for $\delta_{\rm ct} \sim \Gamma_{\rm ct}$)
\begin{equation}
\Delta l \leq 2 \, c \, |\beta'_{\rm sh}| \, \Gamma_{\rm ct} \, t'_{\rm rad} . 
\end{equation}
\noindent

\section{Multiple shocks}

The time scale for particles to escape from the region of the shock ensemble can be estimated as
\begin{equation}
t'_{\rm esc} \sim {D^2 \over \kappa} ,
\end{equation}
\noindent
where $D$ is the thickness of the considered region and $\kappa \sim \lambda_{\rm e}(\gamma) \, c / 3$ is the diffusion coefficient of the electrons with mean free path $\lambda_{\rm e}(\gamma)$ and Lorentz factor $\gamma$. Assuming that $\lambda_{\rm e}(\gamma)$ is comparable to the electron gyroradius,
\begin{equation}
\lambda_{\rm e}(\gamma) \sim r_{\rm g} = {\gamma \, m_{\rm e} c^2 \over e \, B} ,
\end{equation}
\noindent
one obtains
\begin{equation}
{t'_{\rm esc} \over t'_{\rm rad}} \sim {e \, \sigma_{\rm T} \over 2 \pi \, (m_{\rm e} c^2)^2} \, D^2 \, B^3 \, (1+X)
\end{equation}
\noindent
where the radiative loss time scale $t'_{\rm rad}$ is given in Appendix D.

The time scale between subsequent shocks can be estimated as
\begin{equation}
t'_{\rm b} \sim {d' \over c}
\end{equation}
\noindent
where $d'$ is the mean separation of the shocks in the comoving frame of the radiating plasma between the shocks (moving with bulk Lorentz factor $\Gamma \gg 1$). Second-order Fermi acceleration can be characterized by the time scale
\begin{equation}
t'_{\rm F \, II} \sim {\lambda_{\rm e}(\gamma) \over c} \, \left({c \over V_{\rm sc}}\right)^2 ,
\end{equation}
\noindent
where the velocity of the scattering centers involved in the turbulent acceleration $V_{\rm sc}$ can be identified with the Alfven speed, $V_{\rm A}$, and the particle gyroradius can be taken for $\lambda_{\rm e}(\gamma)$ \citep{sta02}. Hence, one obtains the ratio
\begin{equation}
{t'_{\rm b} \over t'_{\rm F \, II}} \sim {d' \over r_{\rm g}} \, \beta_{\rm A}^2 = {e \over m_{\rm e} c^2} \, d' \, B \, \beta_{\rm A}^2 \, \gamma^{-1}
\end{equation}
\noindent
where $\beta_{\rm A} \equiv V_{\rm A} / c$.

\begin{figure}
\plotone{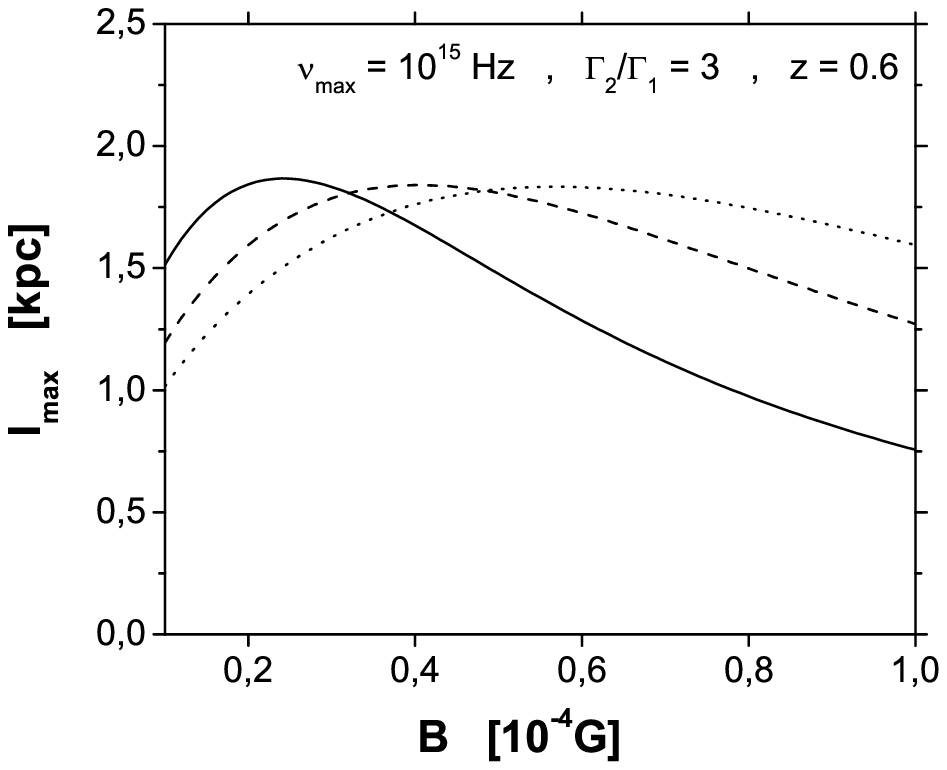}
\caption{ Maximum extent of the emission region in the internal shock model for the synchrotron frequency $\nu_{\rm max} = 10^{15}$ Hz, as a function of the magnetic field $B$ (at redshift $z = 0.6$). The solid line corresponds to the parameters $\Gamma_1 = 3$, $\Gamma_2 = 9$, $\Gamma_{\rm ct} = 5.16$ and $\beta'_{\rm sh} = 0.1826$. The dashed line corresponds to the parameters $\Gamma_1 = 5$, $\Gamma_2 = 15$, $\Gamma_{\rm ct} = 8.64$ and $\beta'_{\rm sh} = 0.18$. The dotted line corresponds to the parameters $\Gamma_1 = 7$, $\Gamma_2 = 21$, $\Gamma_{\rm ct} = 12.11$ and $\beta'_{\rm sh} = 0.1793$. \label{fig1}}
\end{figure}

\begin{figure}
\plotone{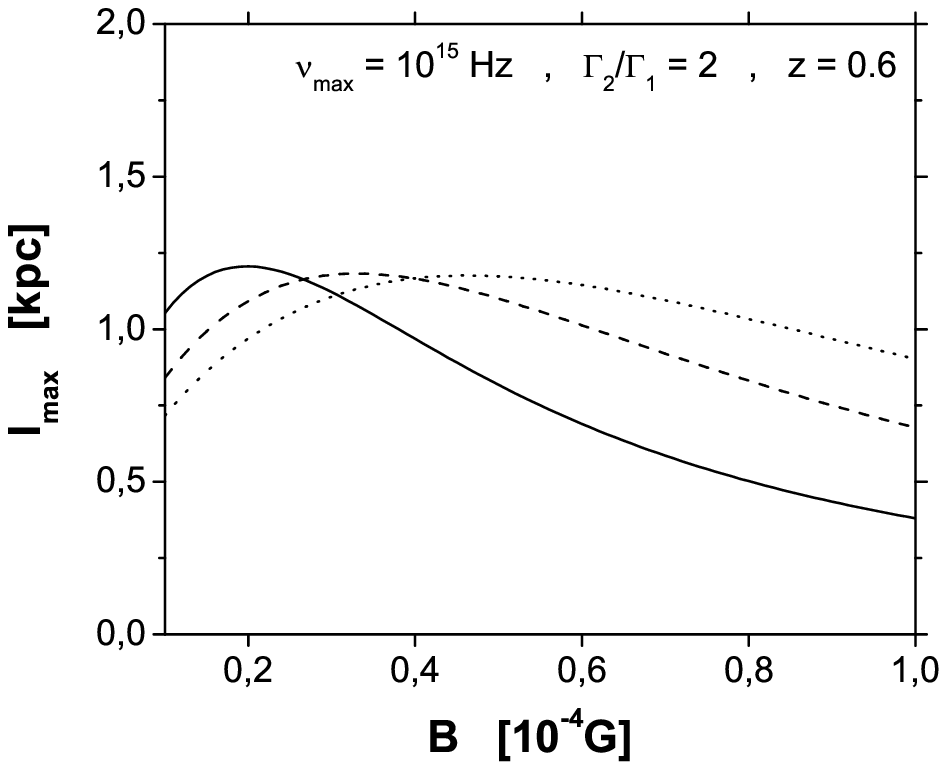}
\caption{ Maximum extent of the emission region in the internal shock model for the synchrotron frequency $\nu_{\rm max} = 10^{15}$ Hz, as a function of the magnetic field $B$ (at redshift $z = 0.6$). The solid line corresponds to the parameters $\Gamma_1 = 3$, $\Gamma_2 = 6$, $\Gamma_{\rm ct} = 4.23$ and $\beta'_{\rm sh} = 0.1179$. The dashed line corresponds to the parameters $\Gamma_1 = 5$, $\Gamma_2 = 10$, $\Gamma_{\rm ct} = 7.06$ and $\beta'_{\rm sh} = 0.116$. The dotted line corresponds to the parameters $\Gamma_1 = 7$, $\Gamma_2 = 14$, $\Gamma_{\rm ct} = 9.89$ and $\beta'_{\rm sh} = 0.115$. \label{fig2}}
\end{figure}

\begin{figure}
\plotone{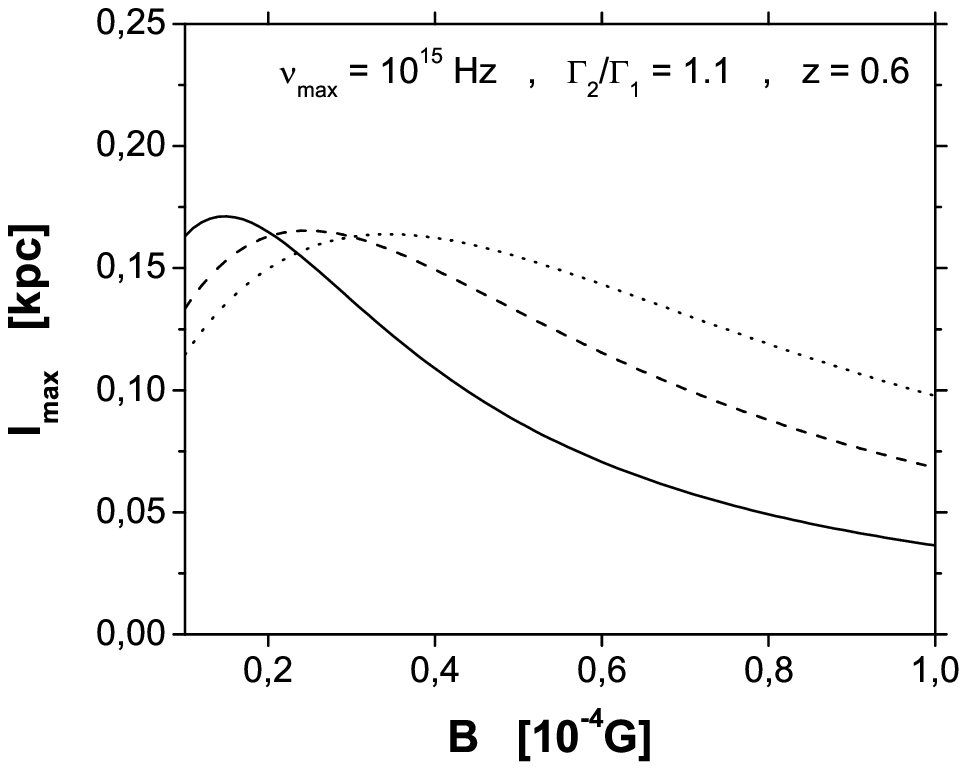}
\caption{ Maximum extent of the emission region in the internal shock model for the synchrotron frequency $\nu_{\rm max} = 10^{15}$ Hz, as a function of the magnetic field $B$ (at redshift $z = 0.6$). The solid line corresponds to the parameters $\Gamma_1 = 3$, $\Gamma_2 = 3.3$, $\Gamma_{\rm ct} = 3.15$ and $\beta'_{\rm sh} = 0.0167$. The dashed line corresponds to the parameters $\Gamma_1 = 5$, $\Gamma_2 = 5.5$, $\Gamma_{\rm ct} = 5.24$ and $\beta'_{\rm sh} = 0.0162$. The dotted line corresponds to the parameters $\Gamma_1 = 7$, $\Gamma_2 = 7.7$, $\Gamma_{\rm ct} = 7.34$ and $\beta'_{\rm sh} = 0.016$. \label{figure}}
\end{figure}

\begin{figure}
\plotone{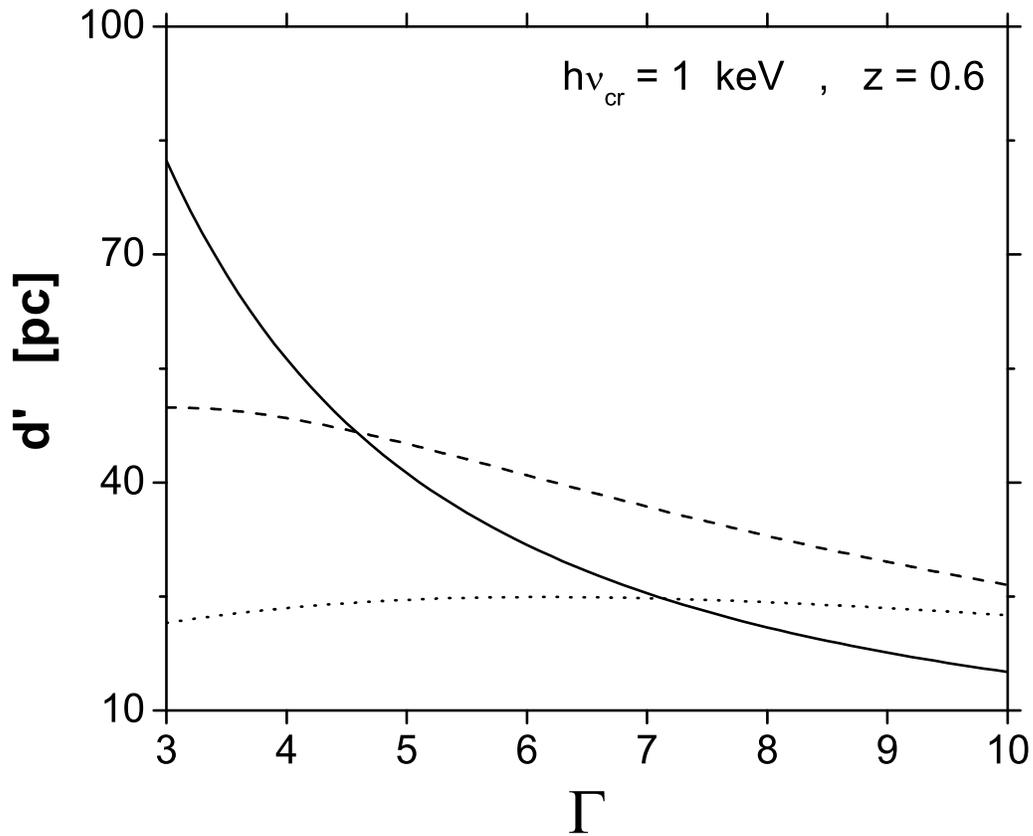}
\caption{ Separation of the shocks in the multiple shock model corresponding to the critical synchrotron photon energy $h \nu_{\rm cr} = 1$ keV, as a function of bulk Lorentz factor $\Gamma$ (for a redshift $z = 0.6$). The solid line corresponds to the magnetic field $B = 10^{-5}$ G, the dashed line to $B = 5 \cdot 10^{-5}$ G and the dotted line to $B = 10^{-4}$ G. \label{fig3}}
\end{figure}

\end{document}